\documentclass[a4paper]{raa_twocolumn}            

\usepackage{graphicx,times}             
\usepackage{natbib}
\usepackage{amssymb,amsmath}
\bibpunct{(}{)}{;}{a}{}{,}

\usepackage[a4paper=true,pagebackref=true,bookmarks=false]{hyperref}
\hypersetup{colorlinks = true, linkcolor = green, anchorcolor = red, citecolor = blue, filecolor = red, pagecolor = red, urlcolor = red}

\def\jref@jnl#1{{\rm#1\/}}
\def\actaa{\jref@jnl{Acta Astronomica}}
\def\aap{\jref@jnl{A\&A}}
\def\aapr{\jref@jnl{The Astronomy and Astrophysics Review}}
\def\aaps{\jref@jnl{Astronomy and Astrophysics Supplement Series}}
\def\aj{\jref@jnl{AJ}}
\def\apj{\jref@jnl{ApJ}}
\def\apjl{\jref@jnl{ApJL}}
\def\apjs{\jref@jnl{ApJS}}
\def\apss{\jref@jnl{Astrophysics and Space Science}}
\def\ao{\jref@jnl{Applied Optics}}
\def\araa{\jref@jnl{ARA\&A}}
\def\bain{\jref@jnl{BAN}}
\def\caa{\jref@jnl{Chinese Astronomy and Astrophysics}}
\def\cjaa{\jref@jnl{Chinese Journal of Astronomy and Astrophysics}}
\def\gca{\jref@jnl{Geochimica et Cosmochimica Acta}}
\def\jcp{\jref@jnl{Journal of Chemical Physics}}
\def\jqsrt{\jref@jnl{Journal of Quantitative Spectroscopy and Radiative Transfer}}
\def\mnras{\jref@jnl{MNRAS}}
\def\memras{\jref@jnl{Memoirs of the Royal Astronomical Society}}
\def\memsai{\jref@jnl{Memorie della Societa Astronomica Italiana}}
\def\na{\jref@jnl{New Astronomy}}
\def\nar{\jref@jnl{New Astronomy Reviews}}
\def\nat{\jref@jnl{Nature}}
\def\pasa{\jref@jnl{Publications of the Astronomical Society of Australia}}
\def\planss{\jref@jnl{Planetary and Space Science}}
\def\pasj{\jref@jnl{Publications of the Astronomical Society of Japan}}
\def\pasp{\jref@jnl{PASP}}
\def\physrep{\jref@jnl{Physics Reports}}
\def\pra{\jref@jnl{Physical Review A}}
\def\prd{\jref@jnl{Physical Review D}}
\def\pre{\jref@jnl{Physical Review E}}
\def\physrep{\jref@jnl{Physics Reports}}
\def\physscr{\jref@jnl{Physica Scripta}}
\def\qjras{\jref@jnl{Quarterly Journal of the Royal Astronomical Society}}
\def\rmxaa{\jref@jnl{Revista Mexicana de Astronomia y Astrofisica}}
\def\skytel{\jref@jnl{Sky and Telescope}}
\def\solphys{\jref@jnl{Solar Physics}}
\def\sovast{\jref@jnl{Soviet Astronomy}}
\def\ssr{\jref@jnl{Space Science Reviews}}
\def\zap{\jref@jnl{Zeitschrift fuer Astrophysik}}
\def\azh{\jref@jnl{Astronomicheskij Zhurnal}}
\def\procspie{\jref@jnl{Proc. SPIE}}

\def\fbs{\textsc{fbs}}
\def\EL{\textsc{ELISa}}

\newcommand{\MC}{\multicolumn}
\newcommand{\kms}{km\,s$^{-1}$}

\newcommand{\Aur}{V454\,Aur}
\newcommand{\NN}{NN\,Del}
\DeclareRobustCommand{\ion}[2]{%
    \relax\ifmmode
    \ifx\testbx\f
    {\mathrm{#1\,\textsc{#2}}}\else
    {\mathrm{#1\,\mathsc{#2}}}\fi
    \else\textup{#1\,{\mdseries\textsc{#2}}}%
    \fi}

\def\fbs{\textsc{FBS}}
\newcommand{\parsec}{\texttt{PARSEC} }
\newcommand{\basti}{\texttt{BaSTI} }
\newcommand{\mist}{\texttt{MIST} }
\newcommand{\deb}{\texttt{DEBCat} }

\begin{document}

    \title{Long-period double-lined eclipsing binaries: the system V454\,Aur with the secondary eclipse caused by the occultation of the hotter component}

   \volnopage{Vol.0 (20xx) No.0, 000--000}      
   \setcounter{page}{1}          

    \author{Alexei Y.\ Kniazev
        \inst{1,2,3}
    \and Oleg Malkov
        \inst{4}
    \and Stanislav Gorda 
        \inst{5}
    \and Leonid N.\ Berdnikov
        \inst{3}
    \and  \\ Ivan Y.\ Katkov
        \inst{6,7,3}
    }

   \institute{%
            South African Astronomical Observatory, Cape Town, 7935, South Africa {\it a.kniazev@saao.nrf.ac.za}\\
        \and
            Southern African Large Telescope, Cape Town, 7935, South Africa\\
        \and
            Sternberg State Astronomical Institute, Moscow, Universitetsky ave., 13, Russia\\
        \and
            Institute of Astronomy, Russian Academy of Sciences, 48 Pyatnitskaya St., Moscow 119017, Russia\\
        \and
            Kourovka Astronomical Observatory, Ural Federal University, Mira st. 19, Yekaterinburg, 620002\\
        \and
            New York University Abu Dhabi, Saadiyat Island, PO Box 129188, Abu Dhabi, UAE\\
        \and
            Center for Astro, Particle, and Planetary Physics, NYU Abu Dhabi, PO Box 129188, Abu Dhabi, UAE
\vs\no
   {\small Received~~2024 December 31; accepted~~20xx~~month day}}

\abstract{We present the results of our study of the long-period eclipsing binary star \Aur. The results are based on spectroscopic data obtained with the UFES \'echelle spectrograph and photometric observations from TESS. The derived radial velocity curve is based on 17 spectra obtained between 2021 and 2023, covering all orbital phases of this binary system. The orbital period determined from TESS data, $P = 27.019803 \pm 0.000003$ days, agrees within uncertainties with the period established in previous studies. The model constructed for the TESS photometric light curve achieves a precision of 0.01\%. The effective temperatures of both components, as well as the system metallicity, were directly derived from the spectra and are $T_\mathrm{eff, A} = 6250 \pm 50$\,K, $T_\mathrm{eff, B} = 5855 \pm 50$\,K, and $\mathrm{[Fe/H]} = -0.10 \pm 0.08$, respectively. Our analysis of the photometric and spectroscopic data allowed us to directly compute the luminosities of the components, $L_A = 1.82\,L_\odot$ and $L_B = 1.07\,L_\odot$, their radii, $R_A = 1.15\,R_\odot$ and $R_B = 1.00\,R_\odot$, and their masses, $M_A = 1.137\,M_\odot$ and $M_B = 1.023\,M_\odot$, with uncertainties below 1\%. Comparison with evolutionary tracks indicates that the system's age is $1.18 \pm 0.10$\,Gyr, and both components are still on the main sequence. The \Aur\ system is particularly interesting due to the partial eclipse of the primary component, which results in the ``inversion'' of the primary and secondary minima in the photometric light curve.
}

   \authorrunning{A.\,Y.\,Kniazev et al.}  
   \titlerunning{Absolute parameters of the V454\,Aur system}  

   \maketitle


\section{Introduction}
\label{txt:intro}

Double-line eclipsing binaries (DLEBs) represent an observational class of binary systems that provides the most accurate (to within 2--3\%) parameters of their components, including such critical yet challenging-to-determine parameters as mass and orbital characteristics. Large-scale studies of DLEBs began approximately half a century ago \citep{1980ARA&A..18..115P,1988BAICz..39..329H,1991A&ARv...3...91A}.  However, current catalogs and lists of such systems still contain only about a hundred objects \citep{2010A&ARv..18...67T,2015ASPC..496..164S}. DLEBs are an indispensable source of data for constructing fundamental relationships for main-sequence stars (e.g., mass-luminosity and mass-radius relations). Consequently, the inclusion of new DLEBs in catalogs is an important and relevant task.

It should be noted that most of the studied DLEB systems are short-period binaries. For example, in the list from \citep{2010A&ARv..18...67T}, only four out of 95 DLEBs have orbital periods exceeding 15 days, and only two of these four ($\alpha$~Cen and AP\,Phe) contain main-sequence (MS) components. In the online catalog \deb\ \citep{2015ASPC..496..164S}, there are currently more than 300 binary systems to date. However, only 150 of these, which have listed luminosities, contain MS components, and just 11 of these systems have orbital periods longer than 15 days.

The study of long-period systems is indeed challenging because it requires extended observation campaigns to construct sufficiently accurate light curves and radial velocity curves. Moreover, the components of such systems exhibit smaller (and thus more difficult to observe) radial velocity variations compared to short-period DLEBs. However, these systems have several advantages over short-period DLEBs. The large separation between components ensures that the DLEB has never been a semi-detached system, with no mass transfer having occurred. Consequently, the current masses of the components are definitively equal to their initial masses (excluding possible mass loss through stellar winds). Additionally, there is a probability that the components of such systems have not undergone significant processes of circularization and synchronization \citep{1975A&A....41..329Z,1977A&A....57..383Z,1988ApJ...324L..71T,1987ApJ...322..856T,2007MNRAS.382..356K,2010MNRAS.401..257K}, at least during their time on the Main Sequence. Thus, the evolution of the components in long-period DLEBs is identical to that of single stars, making them reliable for constructing fundamental stellar relations \citep{2003A&A...402.1055M,2007MNRAS.382.1073M}.

This work continues our series of studies on long-period DLEBs \citep{2020Ap&SS.365..169K,2020RAA....20..119K,2022OAst...31..106P}. Here, we present data on the long-period system \Aur. The variability of the bright (V=7.65 mag) eclipsing binary \Aur\ = HD,44192 was discovered by the Hipparcos satellite, and the first detailed spectroscopic study of the system was conducted by \cite{2001Obs...121..315G}. In that study, radial velocity curves for both components were obtained, and both the orbital elements (period $P=27.0197 \pm 0.0010$ days, eccentricity $e=0.3790 \pm 0.0013$) and the spectral types of the components (F8V + G1/2V) were determined. The component masses were estimated to be $M_A =1.163\,M_\odot$ and $M_B =1.035\,M_\odot$. \cite{2001Obs...121..315G} also evaluated the radii of the components and their rotational velocities, highlighting the possibility of pseudo-synchronization of the components' rotation with the orbital motion, though this was not confirmed. The parallax of \Aur\ obtained by Hipparcos \citep{2007A&A...474..653V} is $14.4\pm0.9$~mas, whereas Gaia \citep{2020yCat.1350....0G} reports a value of $15.367\pm0.022$~mas. Based on data from the Geneva-Copenhagen Survey \citep{2004A&A...418..989N}, the temperature and metallicity of this system were estimated as $T_\mathrm{eff} = 6064$~K and $\mathrm{[Fe/H]} = -0.08$ by \citet{2011A&A...530A.138C}, and as $T_\mathrm{eff} = 6030$~K and $\mathrm{[Fe/H]} = -0.14$ by \citet{2009A&A...501..941H}. 

In the detailed study of \Aur\ by \citet{2024PARep...2...18Y}, the orbital period of the system was determined to be 27.0198177 days, with component masses of $M_A = 1.173\,M_\odot$ and $M_B = 1.045\,M_\odot$, and radii of $R_A = 1.203\,R_\odot$ and $R_B = 0.993\,R_\odot$, respectively. The effective temperatures of the stars were estimated as $T_\mathrm{eff_A} = 6250$~K and $T_\mathrm{eff_B} = 5966$~K. The metallicity of the system was found to be slightly above solar, and the age was estimated to be 1.19\,Gyr. In the detailed study of \Aur\ by \citet{2024Obs...144..181S} the masses and radii of the components were determined as $M_A = 1.161\,M_\odot$, $R_A = 1.211\,R_\odot$ for the primary component, and $M_B = 1.034\,M_\odot$, $R_B = 0.979\,R_\odot$ for the secondary component. The effective temperatures of the stars were estimated as $T_\mathrm{eff_A} = 6170$~K and $T_\mathrm{eff_B} = 5890$~K. In both above detailed studies, the authors used velocity measurements from \cite{2001Obs...121..315G} and photometric data from TESS \citep[Transiting Exoplanet Survey Satellite;][]{2014SPIE.9143E..20R}.

In this study, we also investigate the \Aur\ system using our own spectroscopic \`echelle data and photometric data from the TESS survey.
In Sections \,\ref{txt:phot} and \ref{txt:spec} we describe the available and obtained 
photometric and spectral data, as well as their processing.
In Section \,\ref{txt:data_analysis} the principles of our data analysis are described.
Section \,\ref{txt:res} presents the results we obtained,
they are discussed in Section\,\ref{txt:dis} and summarized in Section\,\ref{txt:sum}.
In the following we will refer to the brighter and hotter star in the DLEB system \Aur\ as component A or primary and the colder star as component B or secondary.

\section{Photometric TESS data}
\label{txt:phot}

For constructing the photometric light curve, we used data from TESS \citep[Transiting Exoplanet Survey Satellite;][]{2014SPIE.9143E..20R}. TESS is a NASA satellite designed to capture images of nearly the entire sky to search for exoplanets using the transit method. The satellite observes a designated area of the sky (referred to as a ``sector'') for a duration of 30 days. Bright stars are observed with a data acquisition cadence of 2 minutes, while full-frame images are recorded every 30 minutes. Due to the satellite's orbit, different regions of the sky are covered non-uniformly; however, there are substantial overlaps between some regions. Most of the processed TESS data are publicly available\footnote{The data used in this work were downloaded from the MAST archive \url{https://mast.stsci.edu}}. In this study, we utilized TESS data processed for the MIT project \citep[``QLP'';][]{2020RNAAS...4..204H,2020RNAAS...4..206H}\footnote{Details are provided at \url{https://archive.stsci.edu/hlsp/qlp}.}. Photometric data for \Aur\ were obtained by TESS during observations of sectors 14, 20, 42--45, and 60, covering both primary and secondary eclipses. The TESS photometric dataset used in this paper contained approximately 48,500 data points, of which only those with a QUALITY flag value of zero were selected.

\begin{table*}
    \centering
    \caption{Radial-velocity observations of \Aur}
    \label{tab:Spec_obs}
    \begin{tabular}{rcccrrrr}
        \hline\hline\\[-0.25cm]
        \# & Date     &   BJD         &  Exp.time     &            \MC{2}{c}{V$_{hel}$}              &            \MC{2}{c}{V$_{model}$}     \\
        &          &               &               &\MC{1}{c}{Star A}       &\MC{1}{c}{Star B}    &\MC{1}{c}{Star A}  &\MC{1}{c}{Star B}  \\
        &          &   (day)       &  (s)          &\MC{1}{c}{(\kms)}       &\MC{1}{c}{(\kms)}    &\MC{1}{c}{(\kms)}  &\MC{1}{c}{(\kms)}  \\
        &   (1)    &   (2)         &  (3)          &\MC{1}{c}{(4)}          &\MC{1}{c}{(5)}       &\MC{1}{c}{(4)}     &\MC{1}{c}{(5)}     \\
        \hline\\[-0.25cm]
        1  & 20211102 & 2459521.54190  & 2$\times$1800 &$-$89.635$\pm$0.056     &   12.960$\pm$0.066  &     $-$89.796     &       12.580      \\
        2  & 20211225 & 2459574.35910  & 3$\times$1800 &$-$78.544$\pm$0.104     &    0.022$\pm$0.046  &     $-$78.531     &        0.058      \\
        3  & 20220112 & 2459592.44100  & 3$\times$1800 &$-$22.369$\pm$0.078     &$-$63.100$\pm$0.088  &     $-$22.030     &    $-$62.746      \\
        4  & 20220225 & 2459636.18182  & 3$\times$1800 &$-$21.733$\pm$0.065     &$-$63.000$\pm$0.052  &     $-$21.713     &    $-$63.099      \\
        5  & 20220310 & 2459649.26460  & 3$\times$1800 &$-$34.962$\pm$0.093     &$-$48.524$\pm$0.117  &     $-$34.723     &    $-$48.637      \\
        6  & 20220325 & 2459664.27910  & 3$\times$1800 &$-$12.045$\pm$0.040     &$-$72.999$\pm$0.049  &     $-$12.501     &    $-$73.339      \\
        7  & 20220411 & 2459681.24850  & 3$\times$1800 &$-$67.807$\pm$0.102     &$-$11.437$\pm$0.102  &     $-$67.932     &    $-$11.723      \\
        8  & 20220421 & 2459691.21580  & 3$\times$1800 &$-$13.431$\pm$0.050     &$-$72.686$\pm$0.040  &     $-$13.021     &    $-$72.760      \\
        9  & 20220929 & 2459852.40848  & 3$\times$1800 &$-$20.812$\pm$0.043     &$-$64.114$\pm$0.138  &     $-$20.951     &    $-$63.946      \\
        10 & 20230115 & 2459960.41445  & 3$\times$1800 &$-$21.821$\pm$0.123     &$-$63.205$\pm$0.123  &     $-$21.770     &    $-$63.035      \\
        11 & 20230311 & 2460015.19456  & 3$\times$1800 &$-$14.780$\pm$0.073     &$-$70.978$\pm$0.066  &     $-$14.826     &    $-$70.754      \\
        12 & 20230318 & 2460022.27413  & 3$\times$1800 &$-$13.000$\pm$0.042     &$-$72.000$\pm$0.259  &     $-$13.335     &    $-$72.412      \\
        13 & 20230330 & 2460034.27233  & 3$\times$1800 &$-$83.242$\pm$0.062     &    6.026$\pm$0.046  &     $-$83.894     &        6.019      \\
        14 & 20230402 & 2460037.22354  & 5$\times$1800 &$-$98.998$\pm$0.054     &   21.699$\pm$0.064  &     $-$98.444     &       22.192      \\
        15 & 20230403 & 2460038.22097  & 5$\times$1800 &$-$85.756$\pm$0.077     &    7.976$\pm$0.103  &     $-$85.992     &        8.351      \\
        16 & 20230409 & 2460044.22458  & 5$\times$1800 & $-$6.735$\pm$0.041     &$-$79.969$\pm$0.042  &      $-$6.647     &    $-$79.845      \\
        17 & 20230410 & 2460045.22352  & 5$\times$1800 & $-$6.621$\pm$0.058     &$-$81.270$\pm$0.068  &      $-$6.001     &    $-$80.563      \\
        \hline\hline
    \end{tabular}
\end{table*}

\section{Spectral observations and data reduction}
\label{txt:spec}

Spectroscopic observations of \Aur\ were conducted from November 2021 to April 2023 using the fiber-fed \'echelle spectrometer UFES \citep{2011AstBu..66..355P,2014AstBu..69..497K} mounted on the 1.21m telescope of the Kourovka Astronomical Observatory at Ural Federal University. A CCD-camera ANDOR DZ936N with back-illuminated and fringe suppression technology sensor BEX2-DD (2048 × 2048, 13.5micron) was used during these observations. The spectrograph was operated with a fiber aperture of 20 arcseconds, yielding a resolution of R$\sim$12,400--13,400. Each observation consisted of three or five exposures of 1800 seconds each, which were subsequently median-combined to remove cosmic ray artifacts. In total, 17 observations were obtained, with the dates listed in Table~\ref{tab:Spec_obs}.

Each night of observations included calibrations consisting of 10 images to account for the zero level (BIAS), three flat-field lamp spectra to determine the positions of the spectral \`echelle orders and to correct for the spectral sensitivity variation along each \`echelle order (the so-called ``blaze correction''), and three hollow-cathode lamp spectra (Th+Ar) for wavelength calibration. These spectra were also median-combined to remove cosmic ray artifacts.

For processing the \`echelle data from the UFES spectrograph, a reduction pipeline was developed based on the data reduction system of the HRS \`echelle spectrograph \citep{2016MNRAS.459.3068K,2019AstBu..74..208K} at the SALT (Southern African Large Telescope). The reduction process included the following steps:
(1) A two-dimensional background, consisting of scattered light, was determined and subtracted from each two-dimensional \`echelle spectrum using the algorithm described in \citet{1996AN....317...95S};
(2) The positions of 66 \`echelle orders were identified using flat-field lamp spectra and extracted from the two-dimensional spectra;
(3) One-dimensional \`echelle orders were corrected for uneven brightness distribution along the orders (the blaze effect) by dividing by the extracted orders of the flat-field spectra;
(4) To construct the dispersion curve, an automatic procedure identified approximately 1600 emission lines in the extracted \`echelle orders of the comparison spectrum. These lines were also automatically located in the two-dimensional \`echelle spectra, and a two-dimensional dispersion curve was constructed for each observing night using a third-order polynomial. Only about 480 emission lines were retained for the final solutions, with the remainder discarded based on various criteria by the automatic procedure. The accuracy of the constructed two-dimensional dispersion curve was approximately $\sim$0.007~\AA;
(5) All extracted \`echelle orders were resampled to a uniform wavelength scale;
(6) All \`echelle orders were combined into a single one-dimensional spectrum. The final spectrum covers the wavelength range of 3850–-7550~\AA\ with a reciprocal dispersion of \mbox{0.032~\AA\ pixel$^{-1}$}.

\begin{figure}
    \includegraphics[clip=,angle=0,width=8.0cm]{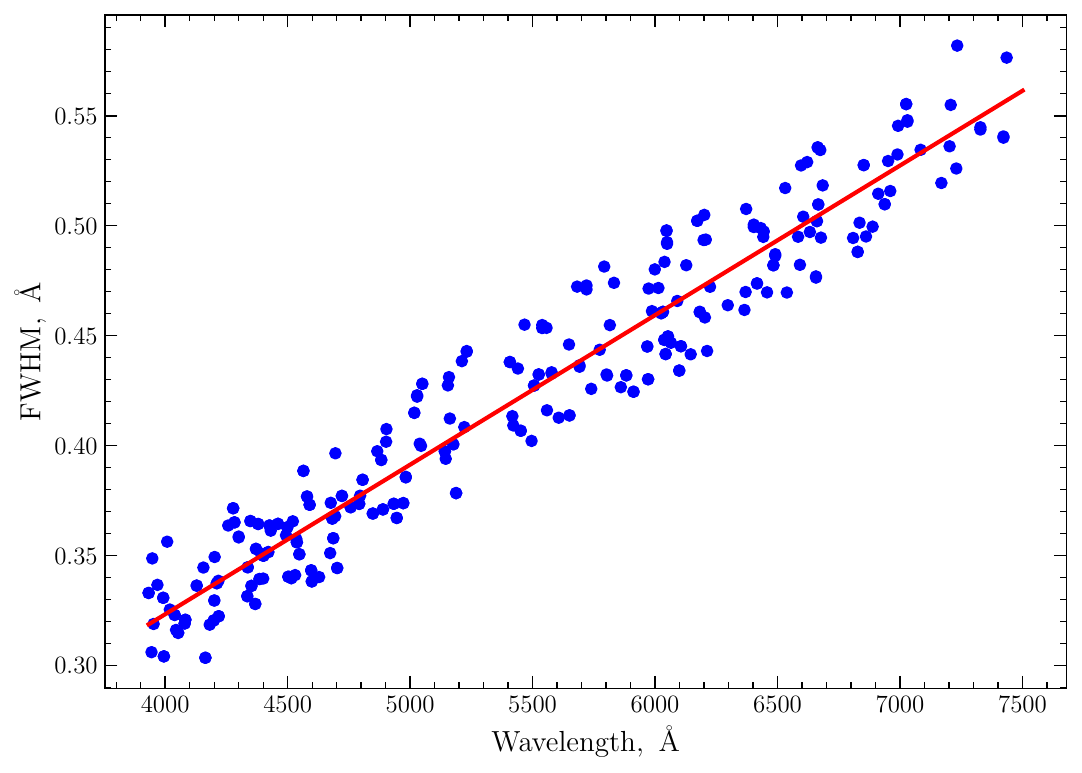}
    \includegraphics[clip=,angle=0,width=8.0cm]{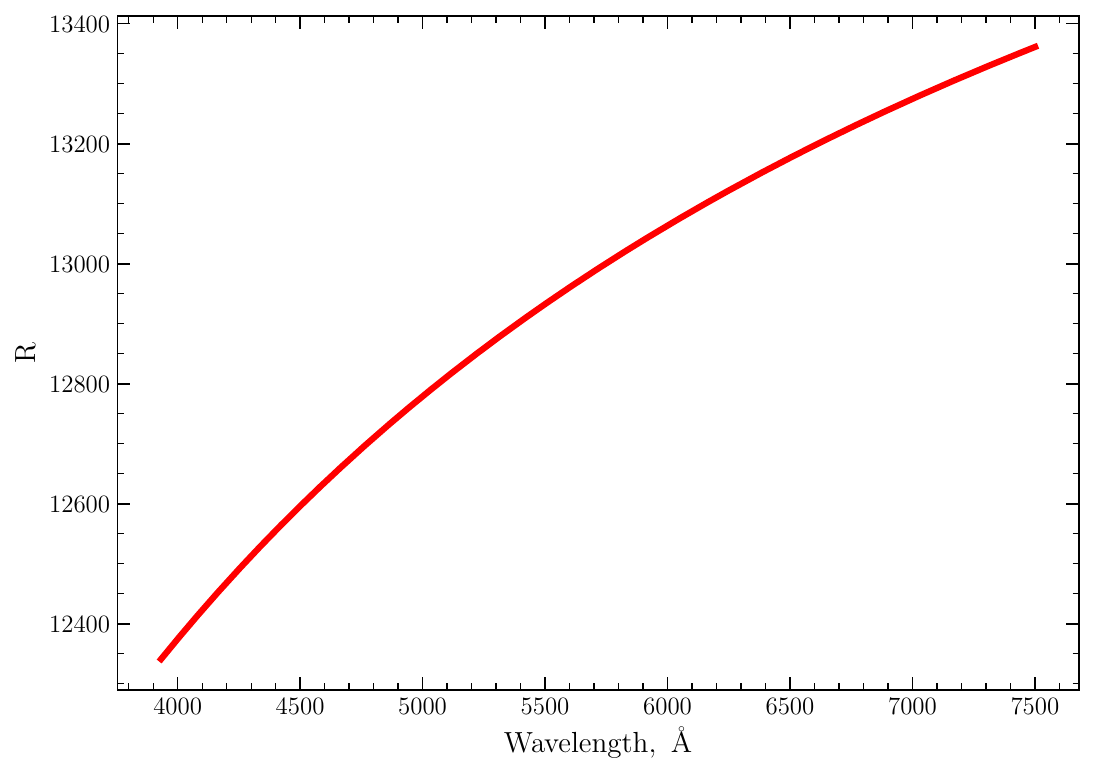}
    \caption{%
        {\bf Top:} Wavelength dependence of spectral resolution (FWHM) for UFES.
        {\bf Bottom:} Resolution value $\rm R = \lambda/\delta\lambda$ as a function of wavelength for UFES.
        \label{fig:FWHM}}
\end{figure}

The spectral resolution (FWHM), measured from all identified lines in the wavelength-calibrated comparison spectrum, varies from $\approx$0.30~\AA\ to $\approx$0.55~\AA\ and is shown in the top panel of Figure~\ref{fig:FWHM}. The behavior of the instrumental profile is well approximated by a first-order polynomial and can be expressed as:
\begin{equation} 
    {\rm FWHM}(\lambda) = 6.80265\cdot10^{-5} \cdot \lambda + 0.05113, 
\end{equation}
for the entire spectral range of 3900--7500~\AA\ with an accuracy of 0.02~\AA. This relationship is also displayed in the top panel of the figure. The resolution $\rm R = \lambda/\delta\lambda$ as a function of wavelength is presented in the bottom panel of Figure~\ref{fig:FWHM}.

\begin{figure*}
    \centering{
        \includegraphics[clip=,angle=0,width=0.75\textwidth]{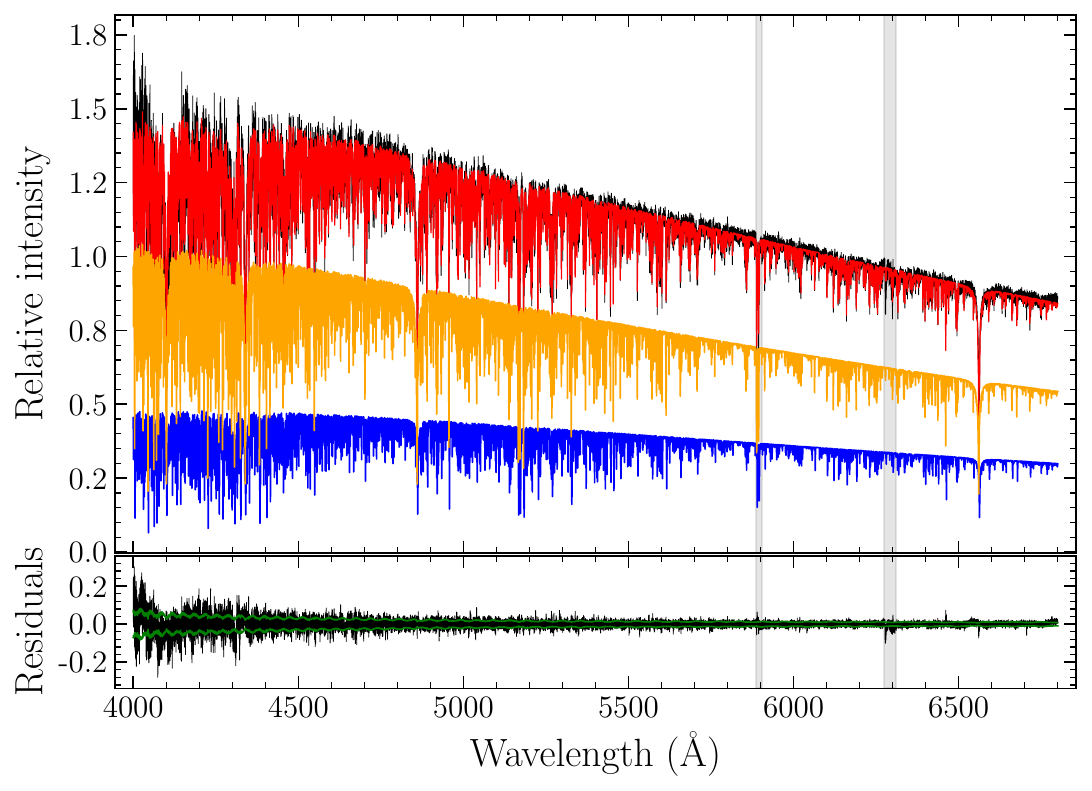}
        \includegraphics[clip=,angle=0,width=0.75\textwidth]{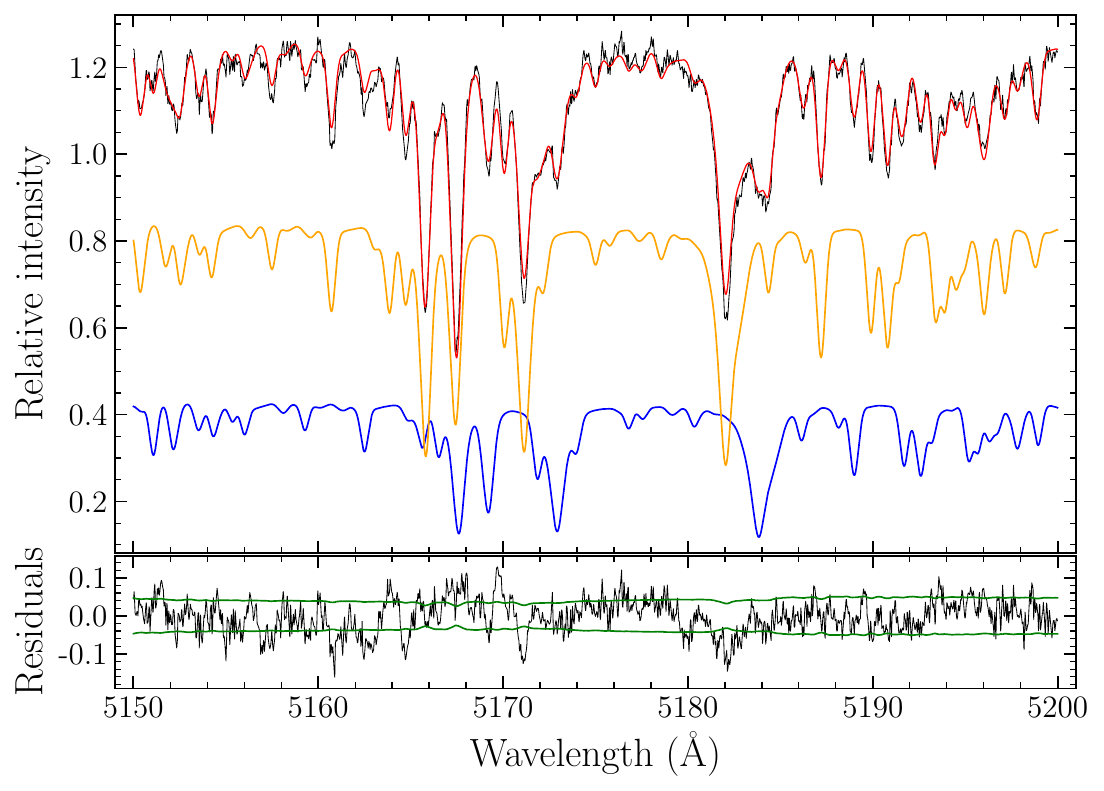}
    }
    \caption{Our analysis of one HRS spectrum of \Aur. The top panel displays the fit of the \'echelle spectrum, with the spectral region around the \ion{Mg}{i} line shown in the bottom panel. Both panels clearly show the observed spectrum in black and the result of the modelling in red. Two components are shown in blue and orange. The bottom panel shows the difference between the observed and modelled spectra in black, with 1$\sigma$\ errors (green lines), which were propagated from the UFES data reduction. Spectral regions excluded from the fit are shown as grey vertical lines.
        \label{fig:spec_fit}}
\end{figure*}

\section{Data analysis}
\label{txt:data_analysis}

To construct the light curves and calculate the period, we used Python programs based on the algorithm from \cite{1965ApJS...11..216L}, which belongs to the class of non-parametric methods and does not require direct application of Fourier decomposition.

For the analysis of fully processed \`echelle spectra, we utilized the \fbs\ package \citep[Fitting Binary Stars;][]{2020RAA....20..119K,2020Ap&SS.365..169K}, specifically developed by our team for the analysis of binary star system spectra. \fbs\ employs a library of theoretically computed high-resolution stellar spectra and is designed to determine radial velocities and stellar parameters ($T_\mathrm{eff}$, $\log g$, $\mathrm{v \sin i}$, [Fe/H]) for both components of a binary system, as well as the parameter $E(B-V)$ for reddening correction and W$_{1,2}$, representing the contribution of each component to the observed spectrum ($\rm W_1 + W_2 = 1$) at the wavelength of the $V$ filter $\lambda$5550~\AA. The program simultaneously fits the observed spectrum with a model spectrum obtained by interpolating the stellar model grid and convolving it with a function that accounts for instrumental resolution and rotational broadening $\mathrm{v \sin i}$, with a shift corresponding to the radial velocity value at a given epoch. For a binary star, the fitting involves modeling the spectra of both components, each with its own radial velocity and stellar atmospheric parameters, effectively decomposing the observed spectrum into the individual spectra of the two components. If multiple spectra of the binary system are available for different epochs, the program can determine a solution in which the parameters ($T_\mathrm{eff}$, $\log g$, $\mathrm{v \sin i}$, [Fe/H], W)${1,2}$ and $E(B-V)$ are consistent across all spectra being fit simultaneously, while the radial velocities of both components, $V{1,2}^j$, are determined for each specific epoch $j$. The stellar models used must be pre-adjusted to match the resolution of the spectrograph in use. The \fbs\ package has already been utilized by the authors for work with both high-resolution and low-resolution spectra \citep[][]{2021MNRAS.503.3856G,2021MNRAS.502.4074M,2022OAst...31..327M,2023MNRAS.523.5510G,2023RAA....23e5021K}.

Additionally, the \fbs\ package allows for the analysis of obtained radial velocities and, by modeling the radial velocity curve, calculates the orbital parameters of the binary system components. During its operation, \fbs\ searches for global minima of sufficiently complex functions \citep{2020RAA....20..119K}. The methods used for finding these minima include various numerical approaches available in the lmfit library\footnote{\url{https://lmfit.github.io/lmfit-py/}}.
In the present study, theoretical stellar models from the library by \citet{Coelho14}, adjusted to match the spectral resolution of UFES (see Section~\ref{txt:spec}), were used. The \fbs\ program was used with only one constraint: it was assumed that both components of the binary system have the same metallicity \citep{2020MNRAS.492.1164H}.

For further analysis of the spectroscopic and photometric data, the \EL\ package \citep[Eclipsing Binary Learning and Interactive System;][]{2021A&A...652A.156C} was used, which constructs a 3D model of the studied system. The \EL\ package enables the determination of absolute parameters for virtually any type of binary system: detached, semi-detached, or contact. To enhance the accuracy of light curve modeling, \EL\ utilizes a library of theoretical stellar spectra, although blackbody radiation models can also be used. \EL\ includes an extensive set of photometric filters, including TESS, various limb darkening laws, and accounts for gravity darkening and reflection effects. For light curve modeling, \EL\ implements Roche geometry and a triangulation process for simulating the surface of binary star components, where the surface parameters of each surface element are treated individually. To generate a point on the light curve, the flux from the entire surface is integrated. \EL\ also provides tools for solving inverse problems, including a built-in Monte Carlo Markov Chain (MCMC) method. This can be applied to determine binary system parameters based on radial velocity datasets or photometric datasets. Additionally, the package supports parallelization for optimization and MCMC computations, significantly reducing computational time when using multi-core computers or processors with many threads.

When calculating the equipotential function, \EL\ uses the synchronization parameter $F_{1,2}$ (for each component), defined as the ratio:
\begin{equation} 
    \label{eqn:synch_det} 
    F = w/w_{orb} 
\end{equation}
where $w$ is the angular rotation velocity of the star, and $w_{orb}$ is the orbital angular velocity of the star.

By default, it is assumed that synchronization occurs rapidly, even in the case of eccentric orbits \citep{1975A&A....41..329Z}. Due to tidal interactions, the synchronization parameter $F = 1$ for circular orbits, whereas for non-circular orbits, synchronization depends on the system's eccentricity and is calculated using the formula from \citet{1981A&A....99..126H}:
\begin{equation} \label{eqn:synch} F = (1 + e)^2/(1 - e^2)^{3/2} \end{equation}
However, in principle, in the case of \EL, the synchronization value $F_{1,2}$ can be a free minimization parameter, and the hypothesis of slow or fast synchronization can be tested in the process of modelling the brightness curve.

When solving the inverse problem, both in the case of radial velocity curve analysis and light curve analysis, \EL\ employs the Least Squares Trust Region Reflective algorithm (LSTRR)\footnote{\url{https://docs.scipy.org/doc/scipy/reference/generated/scipy.optimize.least_squares.html}}. This algorithm is efficient for finding solutions in the local vicinity but does not search for a global minimum. When using the package, it is recommended to provide initial conditions sufficiently close to the true solution\footnote{\url{https://github.com/mikecokina/elisa/blob/dev/ELISa_handbook.pdf}}. For this reason, the orbital parameters of the components in the \Aur\ system and the physical parameters of the stellar components obtained with \fbs\ were used as initial estimates for the \EL\ package.

Thus, the sequence of steps in our analysis was as follows:
(1) Using TESS photometric data, the precise period $P$ and the epoch of the primary minimum $T_0$ were calculated;
(2) Using the \fbs\ package, each observed spectrum was analyzed to obtain the stellar parameters of both components and their velocities;
(3) Using the stellar parameters obtained from the previous step for each spectrum, the mean values and their uncertainties were calculated for the temperatures of each component, their $\log g$, projected rotational velocities, the system metallicity, and the contributions of each component;
(4) Using the \fbs\ package, radial velocity curves were modeled for each component of the system, and initial estimates of the orbital parameters of the binary system were determined;
(5) Using the \EL\ package, radial velocity curves were modeled, and the final orbital parameters of the system were determined, with final uncertainties estimated using the MCMC method;
(6) Using the \EL\ package, TESS photometric data (Step 1), the orbital parameters of the system (Step 5), and the stellar parameters of the components (Step 3), the photometric light curve was modeled, and the absolute parameters of the binary system were determined and their final uncertainties estimated using the MCMC method.

\begin{table}
    \centering
    \caption{Parameters of stellar components for \Aur\ obtained with \fbs\ software}
    \begin{tabular}{lrr}
        \hline\hline\\[-0.25cm]
        Parameter                  &    Star A       &    Star B          \\ \hline\\[-0.25cm]
        $T_\mathrm{eff}$ (K)       & 6250$\pm$50     &  5855$\pm$50       \\
        $\log g$                   & 4.36$\pm$0.03   &  4.44$\pm$0.04     \\
        $v \sin i$ (km s$^{-1}$)   & 1.51$\pm$0.84   &  1.55$\pm$0.91     \\
        $[Fe/H]$                   &  \MC{2}{c}{$-$0.17$\pm$0.02}         \\
        Weight (W)                 & 0.66$\pm$0.01   &  0.34$\pm$0.01     \\
        $E(B-V)$ (mag)             & \MC{2}{c}{0.00$\pm$0.02}             \\
        \hline
    \end{tabular}
    \label{tab:FBS_stars}
\end{table}

\begin{figure*}
    \centering{
        \includegraphics[clip=,angle=0,width=0.70\textwidth]{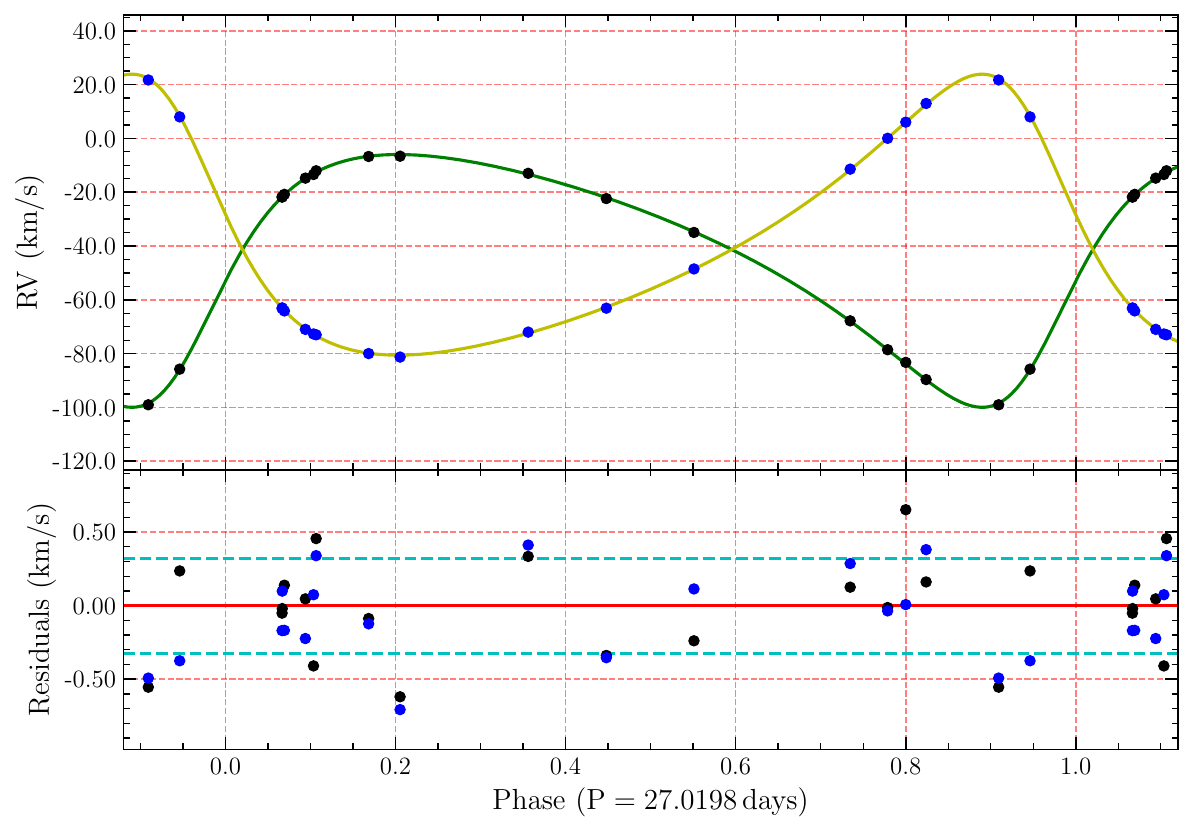}
        \includegraphics[clip=,angle=0,width=0.70\textwidth]{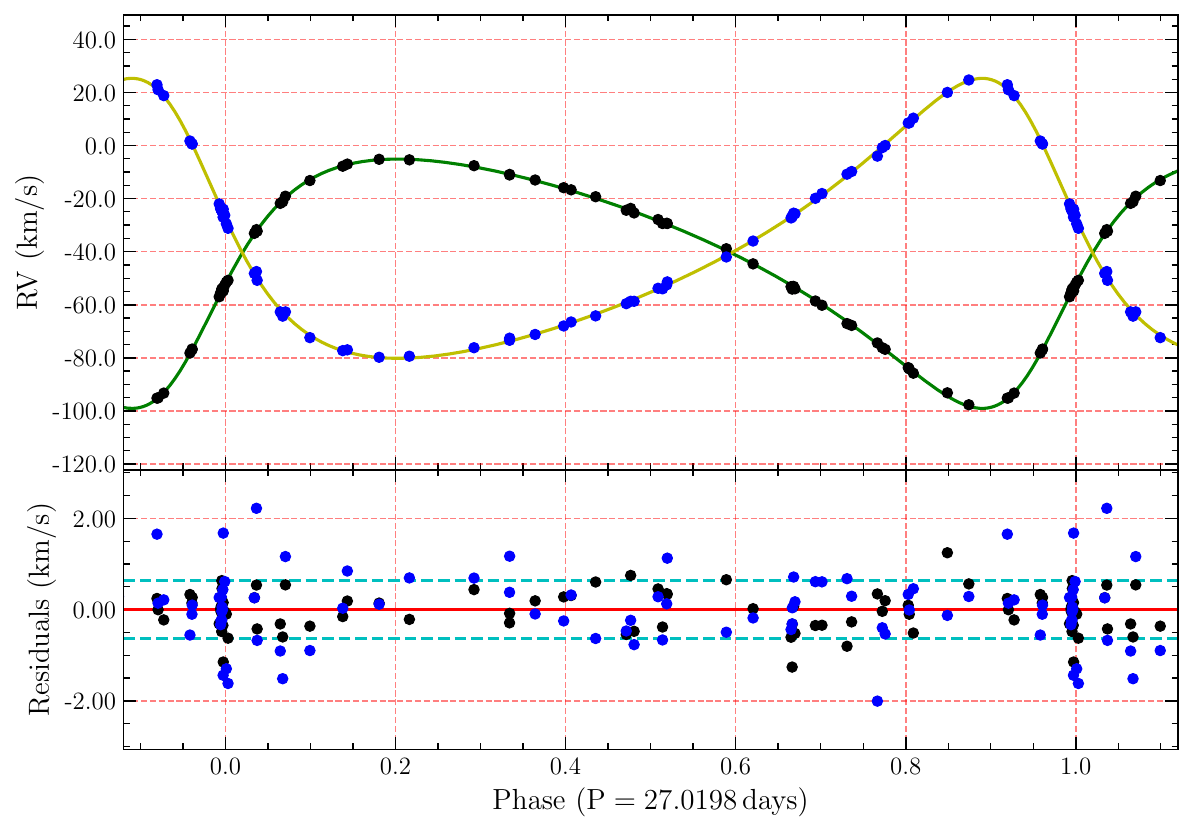}
    }
    \caption{{\it Upper panel:} Radial velocity curve of \Aur\ overlayed with a model fit of \fbs\ software
        with a period of 27.019803~d to our heliocentric radial velocities listed in Table~\ref{tab:Spec_obs}
        and fit parameters are shown in Table~\ref{tab:Orb}. 
        The velocity curve for the primary component is shown in green, 
        for the secondary component in yellow. Phase 0.0 is aligned with the primary minimum for the brightness curve.
        Residuals of the fit, with a rms of 0.32~\kms\ are shown by the light blue dashed line.
        {\it Bottom panel:} Velocity curve for the \Aur\ system, recalculated with the package \fbs,  
        for points from the \citet{2001Obs...121...315G}.
        Residuals of the fit, with a rms of 0.63~\kms\ are shown by the light blue dashed line.
        \label{fig:FBS_fitV}}
\end{figure*}

\begin{table*}
    \centering
    \caption{Orbital parameters of the binary system \Aur}
    \begin{tabular}{llll}
        \hline\hline\\[-0.25cm]
        Parameter                                                   &       \fbs\ (our)           &     \EL\ (our)                   &      \fbs\ (G2001)          \\ \hline\\[-0.25cm]
        Orbital period $P$ (d) (fixed)                              &  27.019803                  & 27.019803                        & 27.019803                   \\
        Epoch at periastron passage $T_p$ (JD)                      &  2439989.6506$\pm$0.0220    & ---                              & 2439989.656$\pm$0.0155      \\
        Epoch at primary minimum $T_0$ (JD)                         &  2439981.8777               & 2439981.8777 (fixed)             & 2439981.8565                \\
        RV semi-amplitude $K1$ (km s$^{-1}$)                        &  46.972$\pm$0.125           & ---                              & 46.983$\pm$0.147            \\
        RV semi-amplitude $K2$ (km s$^{-1}$)                        &  52.212$\pm$0.138           & ---                              & 52.753$\pm$0.148            \\        
        Mass ratio $q=M_2/M_1$                                      &  0.8996$\pm$0.0034          &   0.8996$\pm^{0.0006}_{0.0005}$  & 0.888$\pm$0.005             \\        
        Eccentricity $e$                                            &  0.3806$\pm$0.0018          &   0.3806$\pm^{0.0002}_{0.0002}$  & 0.381$\pm$0.002             \\
        Systemic heliocentric velocity $\gamma$ (km s$^{-1}$)       &$-$41.312$\pm$0.063          &$-$41.314$\pm^{0.011}_{0.010}$    & $-$40.47$\pm$0.06           \\
        The longitude of the periastron $\omega$  (degrees)         &  229.30$\pm$0.33            &   229.30$\pm^{0.03}_{0.03}$      & 229.53$\pm$0.30             \\
        $a*\sin(i)$ ($R_\odot$)                                     &  48.99$\pm$0.10             &    49.98$\pm^{0.03}_{0.02}$      & 49.236$\pm$0.150            \\ 
        Residuals of Keplerian fit (km s$^{-1}$)                    &  0.323                      & ---                              & 0.632                       \\
        \hline
    \end{tabular}
    \label{tab:Orb}
\end{table*}

\begin{figure*}
    \centering{
        \includegraphics[clip=,angle=0,width=15.5cm]{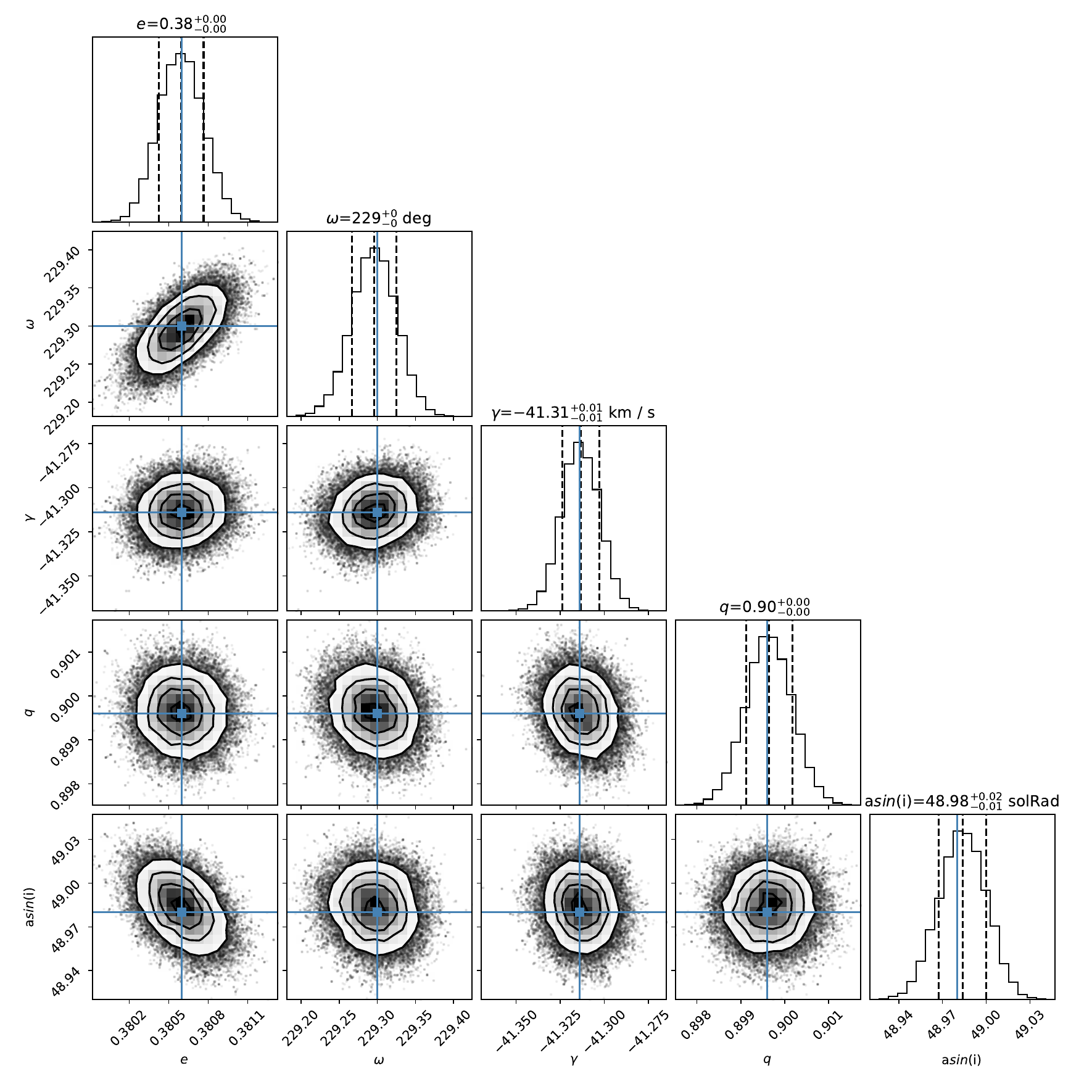}
    }
    \caption{The corner plot of the posterior distributions for the determined orbital parameters shows that most parameters exhibit weak correlations with each other, with the exception of the pairs $e$ -- $\omega$ and $e$ -- $a \sin(i)$, which have moderately low correlation coefficients. To construct these distributions and estimate uncertainties, 400,000 MCMC models were generated.
        \label{fig:V454_RV_corner}}
\end{figure*}

\section{Results}
\label{txt:res}

(1) Using TESS data and a program based on the algorithm from \cite{1965ApJS...11..216L}, the orbital period was determined to be $P = 27.019803 \pm 0.000003$ days, and the epoch of the primary minimum in the system was calculated as BJD $T_0 = 2458850.801464 \pm 0.000105$.

(2) The result of analyzing one spectrum of \Aur\ is shown as an example in Figure~\ref{fig:spec_fit}. The top panel of the figure displays the modeling result in the spectral range of 3900--6800~\AA, while the bottom panel shows the modeling result for the region around the \ion{Mg}{i} line. In each panel, the black and red lines correspond to the observed spectrum and its modeled fit, respectively. The blue and orange spectra represent the modeled spectra of components A and B, respectively. The lower part of each panel shows the difference between the observed and modeled spectra, including the uncertainties obtained during the data processing, which are shown in green.

(3) Experience with \fbs\ shows that the most robust and accurate results in terms of finding the global minimum are obtained using the ``differential evolution'' method. Unfortunately, the time required to find the global minimum with this method, even for a single \`echelle spectrum, is substantial and increases nonlinearly when working simultaneously with more than one spectrum. This issue also prevents error estimation via the Monte Carlo method. For this reason, a ``pseudo''-statistical approach was used to determine the stellar parameters. In this approach, all observed spectra were paired iteratively, and the pairs were processed by the program. With 17 observed spectra, 136 pairs were analyzed. For each solution, the mass ratio $q$ was calculated using the formula:
\begin{equation} 
    q = -(V_{1} - \gamma)/(V_{2} - \gamma), 
\end{equation}
where $\gamma$ is the systemic heliocentric velocity of the \Aur\ system, and $V_1$ and $V_2$ are the measured velocities of each star for each observation pair. Unstable solutions that deviated significantly from the mean value of $q$ were discarded (approximately 30\% of the solutions) through iterative filtering at the $2.5\sigma$ level. The remaining solutions were used to calculate the means and their uncertainties for the parameters of each component: temperatures $T_\mathrm{eff}$, $\log g$, projected rotational velocities $\mathrm{v \sin i}$, metallicity [Fe/H], and contributions to the flux at a wavelength of 5500~\AA. The physical parameters of the stellar components and their uncertainties determined in this way are presented in Table~\ref{tab:FBS_stars}.

(4) The calculated barycentric velocities for both components of the \Aur\ system at specific epochs, along with their associated uncertainties, are presented in Table~\ref{tab:Spec_obs}. The results of modeling these velocities with the \fbs\ program, in the form of calculated radial velocity curves as a function of the observational phase, are shown in Figure~\ref{fig:FBS_fitV}. The derived orbital parameters of the \Aur\ system components and their uncertainties are listed in Table~\ref{tab:Orb}. We also utilized published velocities from \citet{2001Obs...121..315G} to calculate the orbital parameters of the \Aur\ components using the \fbs\ program and compared them with the parameters obtained from our observations. The results of modeling these velocities with the \fbs\ program are also shown in Figure~\ref{fig:FBS_fitV}, and the derived orbital parameters of the \Aur\ components are listed in Table~\ref{tab:Orb}. The comparison shows that the orbital parameters of the system agree well within the uncertainties. However, the data from \citet{2001Obs...121..315G} exhibit twice the scatter compared to our data, despite having a significantly larger number of observational points -- 52 measurements versus 17. The epoch of the primary minimum also agrees very well within the uncertainties, indicating that no significant apsidal motion is observed in the spectral data over the past 22 years.

(5) As the first step in working with \EL, the radial velocity curve data provided in Table~\ref{tab:Spec_obs} were analyzed. Unlike the \fbs\ package, \EL\ uses the mass ratio $q = M_2/M_1$ and the parameter $a \sin(i)$ as input and output parameters. Initial values for these parameters were taken from the results of the \fbs\ analysis, as shown in the corresponding column. The results of \EL\ for modeling radial velocity curves and estimating parameter uncertainties based on posterior distributions using MCMC are also presented in Table~\ref{tab:Orb}. To ensure good statistical reliability in evaluating parameter values and their uncertainties, 400,000 solutions were generated with MCMC. The first 100,000 generations were discarded to eliminate autocorrelation effects, and the remaining generations were used to calculate mean values and confidence intervals. It is evident that the orbital parameter values from both programs agree within the uncertainties. The corner plot of the posterior distributions for the determined orbital parameters is shown in Figure~\ref{fig:V454_RV_corner}. It reveals that most parameters exhibit weak correlations with each other, with the exception of the pairs $e$ -- $\omega$ and $e$ -- $a \sin(i)$, which show moderate correlation coefficients of approximately 0.5 and $-0.5$, respectively.

(6) Using the derived orbital parameters, the calculated mass ratio $q = M_2/M_1$, and the physical parameters of the components such as metallicity, $T_\mathrm{eff}$, and $\log g$, the TESS light curve was analyzed. The temperatures of the components and the metallicity were considered fixed and taken from Table~\ref{tab:FBS_stars}, with the components assumed to be synchronized in the sense of Equation~\ref{eqn:synch}. The variable parameters included the surface potential of each component $\Omega_{1,2}$, the inclination angle of the system $i$, as well as the system's eccentricity $e$ and longitude of periastron $\omega$. While the values of $e$ and $\omega$ were already known from modeling the radial velocity curves, the high-precision TESS light curve allowed for their refinement, despite the known correlation between these parameters, as illustrated in Figure~\ref{fig:V454_RV_corner}.

\begin{figure}
    \centering{
        \includegraphics[clip=,angle=0,width=0.5\textwidth]{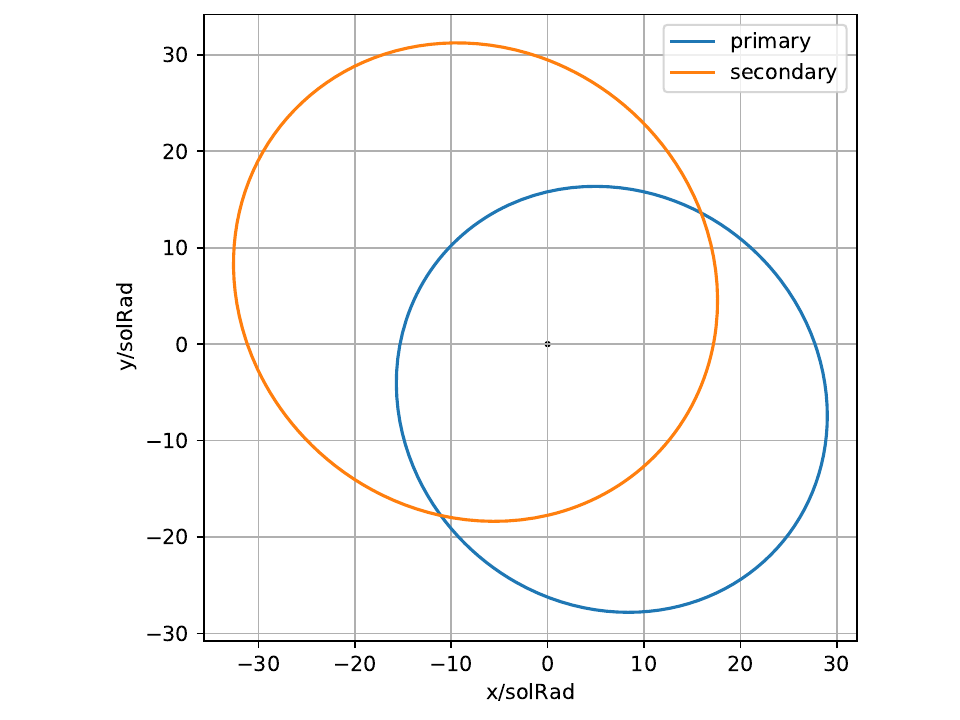}
    }
    \caption{Orbits of the primary and secondary components of the \Aur\ system in the baryocentric coordinate system. The observer looks at the system from the left and phase 0 corresponds to the location of both components on the Y=0 axis.
        \label{fig:V454_orbit}}
\end{figure}

\begin{figure}
    \centering{
        \includegraphics[clip=,angle=0,width=0.45\textwidth]{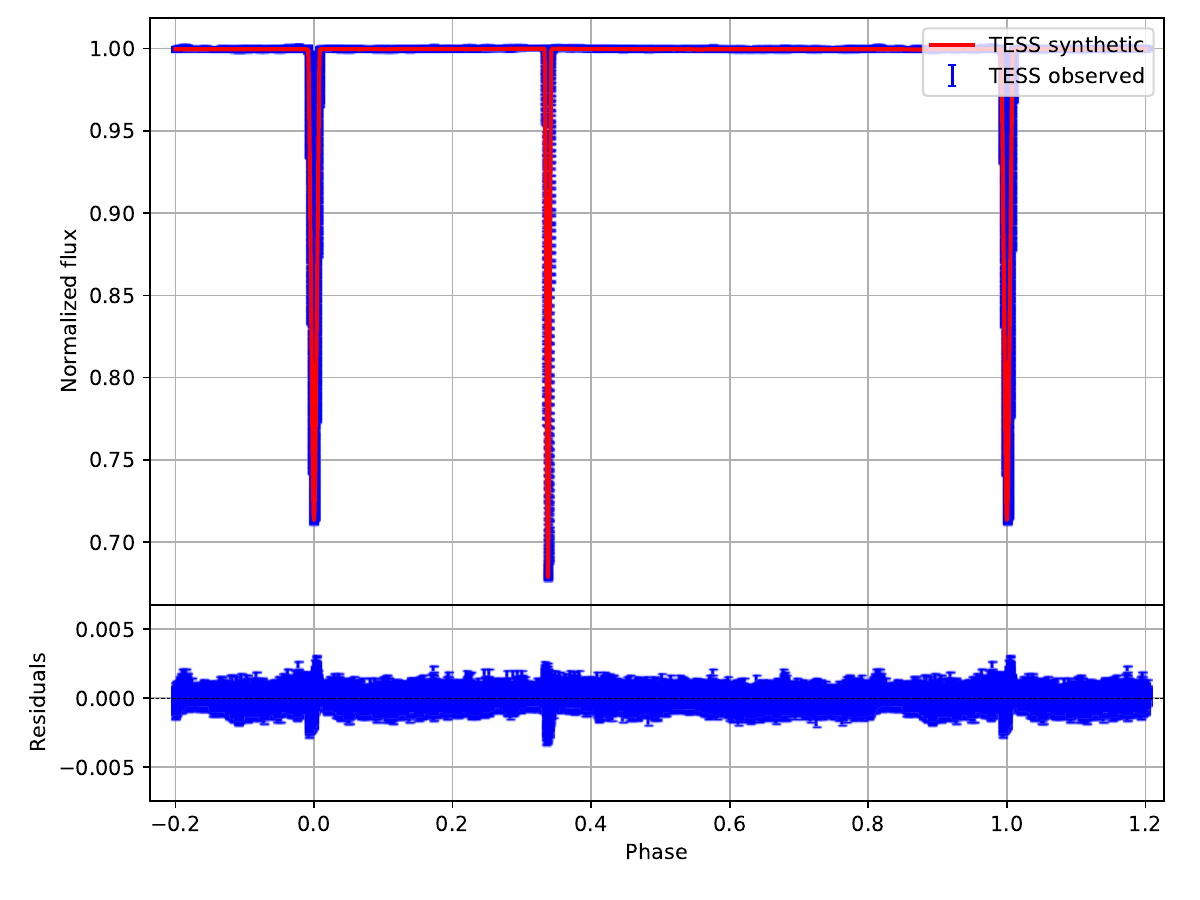}
        \includegraphics[clip=,angle=0,width=0.45\textwidth]{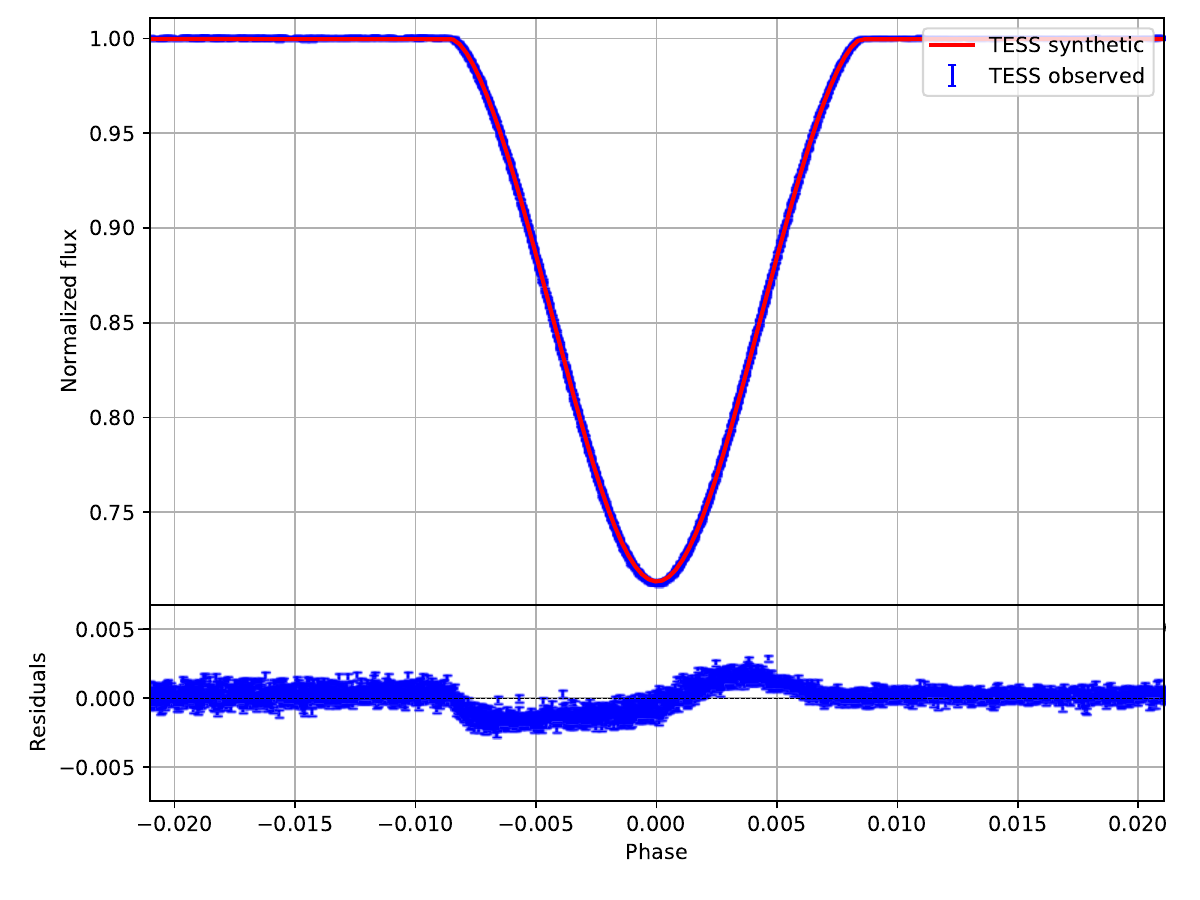}
        \includegraphics[clip=,angle=0,width=0.45\textwidth]{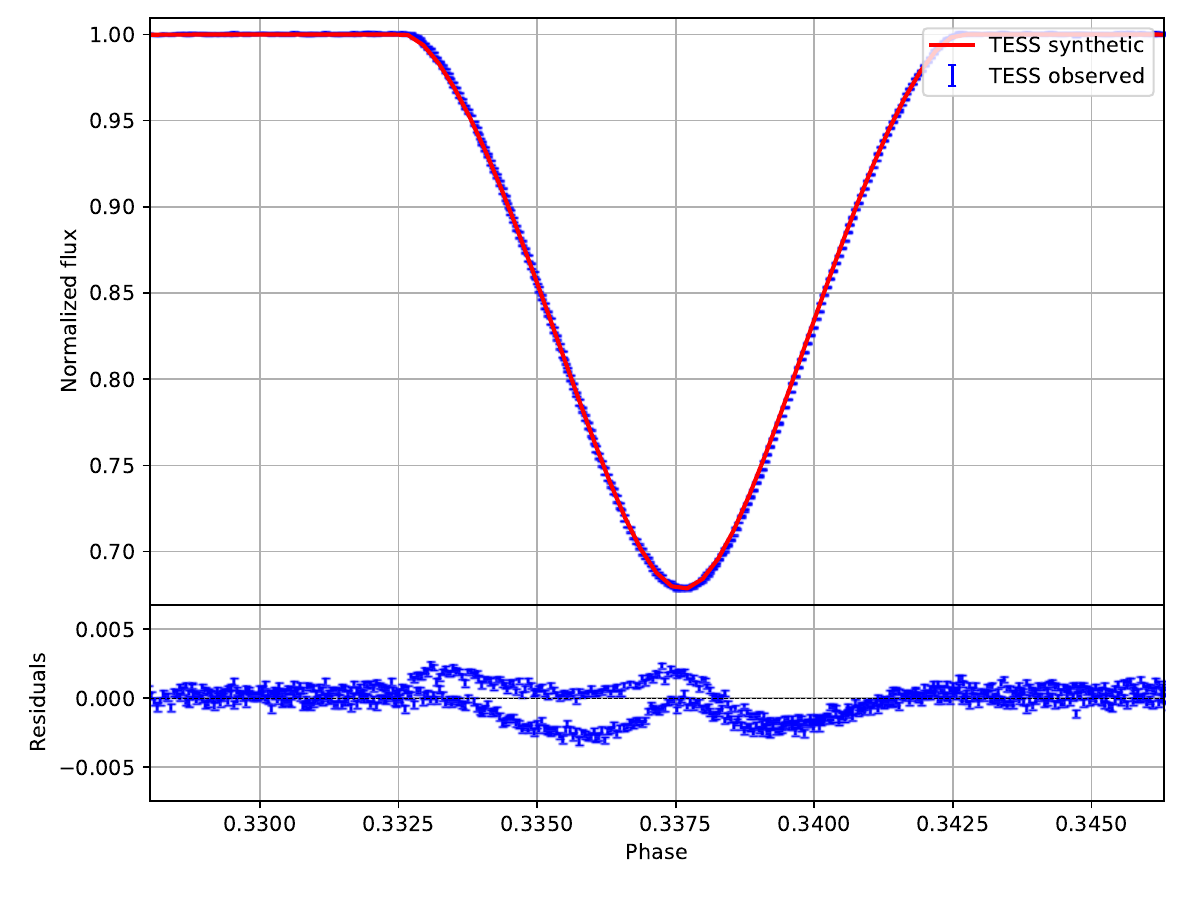}
    }
    \caption{The brightness curve from TESS for \Aur\ fitted with \EL. The eclipse of the hotter component A of the \Aur\ system is ``more partial'' than the eclipse of the cooler component B and for that reason the ``true'' primary minimum, corresponding to the hotter component, is not as deep as the secondary minimum.
        \label{fig:phot_TESS_elisa}}
\end{figure}

The orbits of the components of the \Aur\ system are shown in Figure~\ref{fig:V454_orbit}, and the brightness curve is displayed in Figure~\ref{fig:phot_TESS_elisa}. In Figure~\ref{fig:V454_orbit}, the observer views the system from the left, and phase 0 (eclipse of the hotter component) corresponds to the alignment of both components along the Y-axis at $Y=0$. Modeling of the brightness curve reveals that the eclipse of the hotter component A of the \Aur\ system is ``more partial'' than the eclipse of the cooler component B. As a result, the ``true'' primary minimum, corresponding to the hotter component, is not as deep as the secondary minimum. Consequently, the minima appear ``switched'' on the brightness curve, as shown in Figure~\ref{fig:phot_TESS_elisa} and in Figure~\ref{fig:V454_orbit}. Once this fact was understood, further modeling posed no difficulties. Using only five variables ($\Omega_{1,2}$, $i$, $e$, $\omega$), we were able to construct a model with a precision of approximately 0.1\% ($\chi^2 = 0.999$). The use of various limb darkening laws available in \EL\ led to differences in the model fit only in the fifth decimal place. Ultimately, we chose the ``square root'' law, which performed slightly better than the linear or logarithmic options. The reflection parameter (albedo) was used with its default value, as it did not affect the model quality, which is reasonable in this case.

\begin{figure*}
    \centering{
        \includegraphics[clip=,angle=0,width=16.0cm]{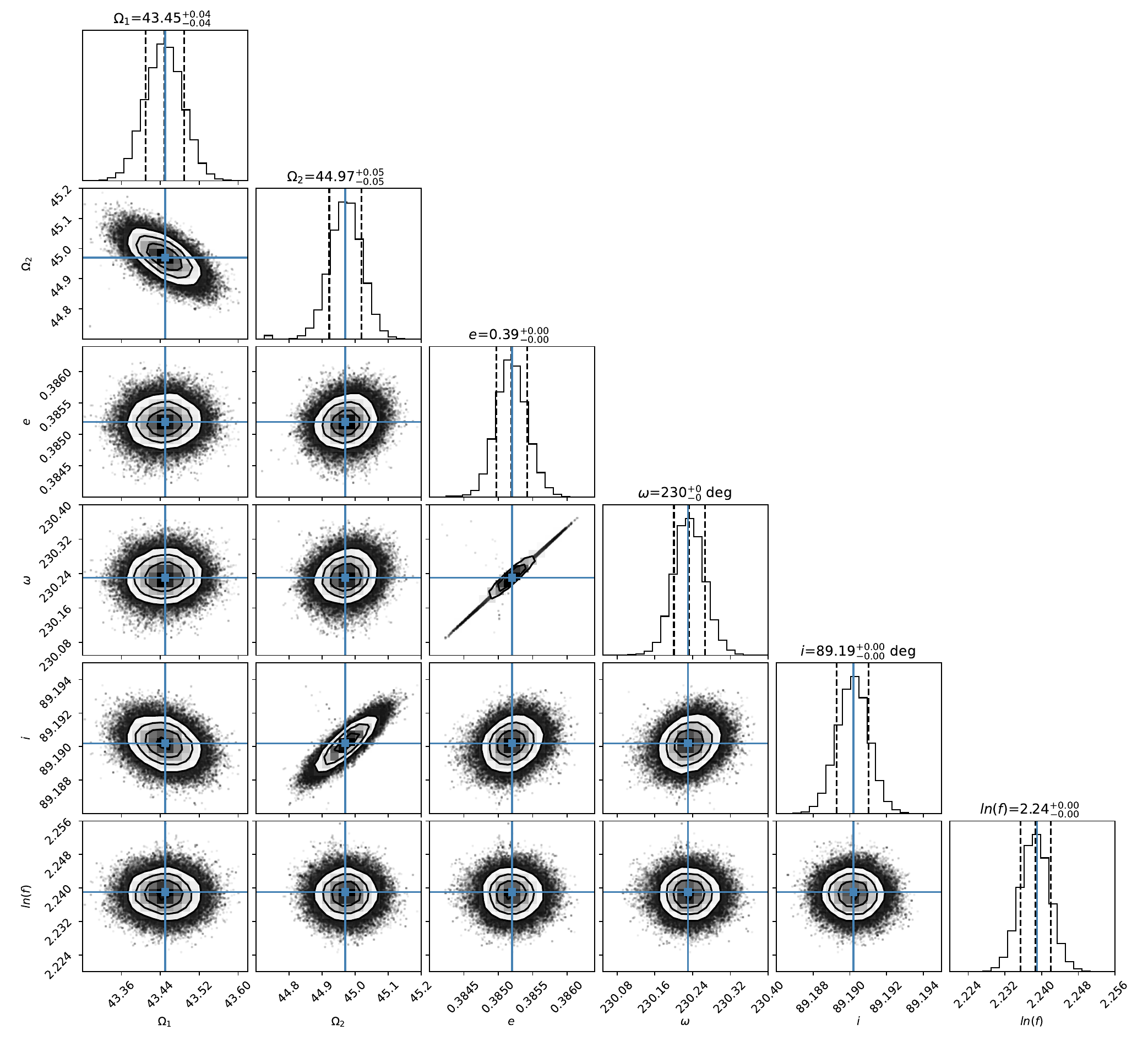}
    }
    \caption{The corner plot of the posterior distributions for the determined absolute parameters of the system. To improve the reliability of the confidence intervals, a robust parameter $\ln(f)$ is used.
        \label{fig:V454_LC_corner}}
\end{figure*}

Subsequently, uncertainties were estimated based on posterior distributions constructed using the MCMC method with 400,000 model generations. The first 150,000 generations were discarded and the remaining generations were used to calculate mean values and confidence intervals. The final results and their confidence intervals are presented in Table~\ref{tab:glob_ELSA}. The resulting model, based on these parameter values, is shown in Figure~\ref{fig:phot_TESS_elisa}, and the posterior distributions of the fitted parameters ($\Omega_{1,2}$, $i$, $e$, $\omega$) are displayed in Figure~\ref{fig:V454_LC_corner}.



\begin{table*}
    \centering
    \caption{Absolute parameters of \Aur}
    \begin{tabular}{lrr}
        \hline\hline\\[-0.25cm]
        Parameter                                           &  \MC{2}{c}{Value}                             \\ \hline\\[-0.25cm]      
        Inclination $i$ (degrees)                           &  \MC{2}{c}{$\,\,\,\,\,89.190\pm0.001$}        \\ 
        Semi-major axis (SMA) $a$ ($R_\odot$)               &  \MC{2}{c}{$48.9849\pm0.0001$}                \\ 
        Eccentricity $e$                                    &  \MC{2}{c}{$\,0.3852\pm0.0002$}               \\
        The longitude of the periastron $\omega$  (degrees) &  \MC{2}{c}{$\,230.232\pm0.032$}               \\ \hline\\[-0.25cm]
        & \MC{1}{c}{Component A} & \MC{1}{c}{Component B}\\ \hline\\[-0.25cm]      
        Mass $M$ ($M_\odot$)                          &  1.1373$\pm$0.0001     &  1.0231$\pm$0.0001         \\
        Surface potential $\Omega$                    &  43.45$\pm$0.04        &  44.97$\pm$0.05            \\
        Synchronicity $F$                             &  2.4414$\pm$0.0015     &  2.4414$\pm$0.0016         \\
        log g (dex)                                   &  4.372$\pm$0.001       &  4.445$\pm$0.001           \\
        Equivalent radius $R_{equiv}$  (SMA)          &0.023494$\pm$0.00022    &0.020475$\pm$0.00022        \\
        Radius $R$ ($R_\odot$)                        &  1.1509$\pm$0.0108     &  1.0030$\pm$0.0108         \\
        Bolometric luminosity $L_{bol}$ ($L_\odot$)   & 1.823$\pm$0.003        &  1.065$\pm$0.002           \\
        \hline\\[-0.25cm]
    \end{tabular}
    \label{tab:glob_ELSA}
\end{table*}

\begin{figure}
    \centering{
        \includegraphics[clip=,angle=0, width=0.450\textwidth]{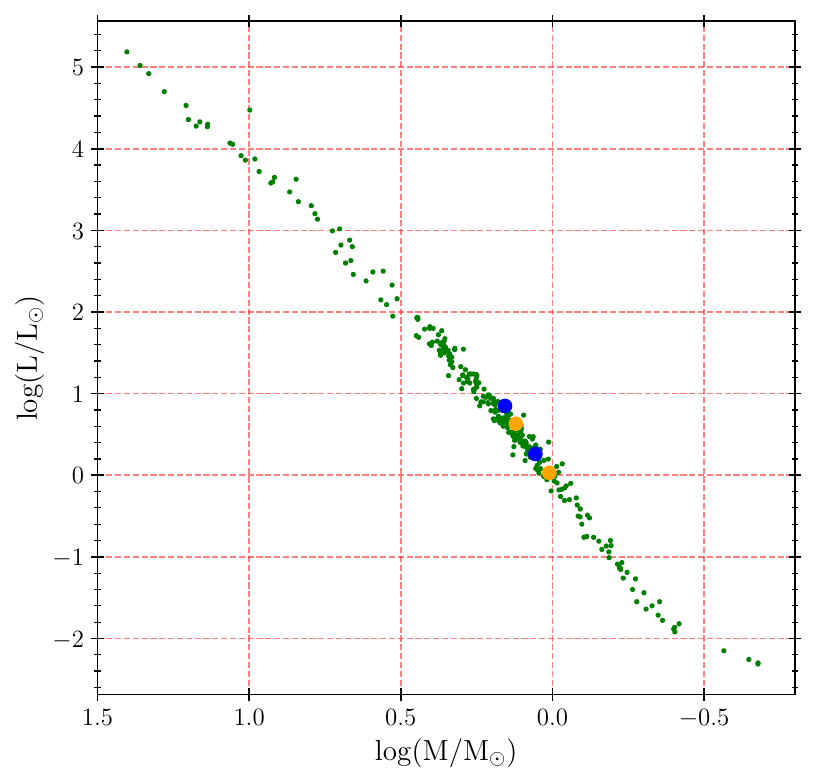}
        \includegraphics[clip=,angle=0, width=0.450\textwidth]{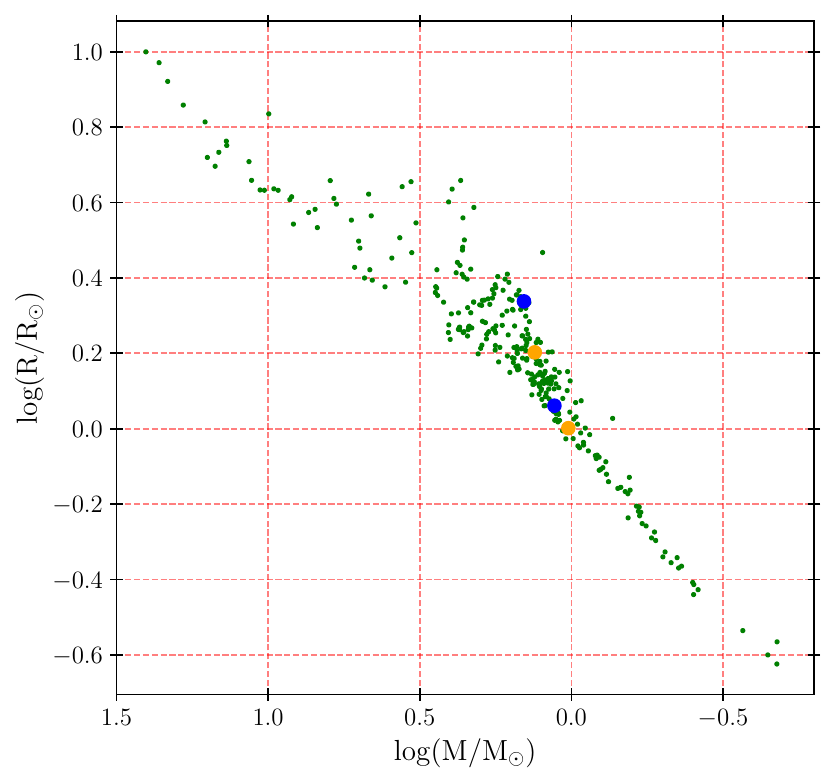}
    }
    \caption{Mass-luminosity (top) and mass-radius (bottom) diagrams for all stars from \citet{2015ASPC..496..164S} that are on the main sequence and have luminosity estimates. The positions of both components of \Aur\ from this work and the \NN\ system \citep{2020Ap&SS.365..169K} ($P=99.25$~days) are shown on both diagrams. Component A is marked in blue, while component B is marked in orange. The symbol sizes are significantly larger than the uncertainties in the values.
        \label{fig:our_MLR}}
\end{figure}

\section{Discussion}
\label{txt:dis}

\begin{table*}
    \centering
    \begin{tabular}{cccc}
        \hline\hline\\[-0.25cm]
        \textbf{Parameter}    & \citet{2024PARep...2...18Y}     & \citet{2024Obs...144..181S} &     This work     \\ \hline\\[-0.25cm]
        $M_A$, $M_\odot$      & $1.173 \pm 0.016$               & $1.161 \pm 0.008$           & $1.1373 \pm 0.0001$     \\
        $M_B$, $M_\odot$      & $1.045 \pm 0.015$               & $1.034 \pm 0.006$           & $1.0231 \pm 0.0001$     \\
        \hline\\[-0.25cm]                                                                                                          
        $R_A$, $R_\odot$      & $1.203 \pm 0.022$               & $1.211 \pm 0.003$           & $1.1509 \pm 0.0108$     \\
        $R_B$, $R_\odot$      & $0.993 \pm 0.034$               & $0.979 \pm 0.003$           & $1.0030 \pm 0.0108$     \\
        \hline\\[-0.25cm]                                                                                                          
        $T_\mathrm{eff_A}$, K & $6250 \pm 150$                  & $6170 \pm 100$              &   $6250 \pm 50$         \\
        $T_\mathrm{eff_B}$, K & $5966^{+109}_{-89}$             & $5890 \pm 100$              &   $5855 \pm 50$         \\
        \hline\\[-0.25cm]                                                                                                    
        $P$, d                & $27.0198177 \pm 0.0000003$      & --                          & $27.019803 \pm 0.000003$\\
        \hline\\[-0.25cm]                                                                                                    
        $i$, (degrees)        & $89.263^{+0.025}_{-0.027}$      & $89.2084 \pm 0.0023$        & $89.190 \pm 0.001$      \\
        Eccentricity $e$      & $0.37717^{+0.00016}_{-0.00013}$ & $0.38056 \pm 0.00017$       & $0.3852 \pm 0.0002$     \\
        $a$, ($R_\odot$)      & $49.418^{+0.173}_{-0.167}$      & $49.24 \pm 0.10$            & $48.9849\pm0.0001$      \\
        Distance, parsec      & $65^{+2}_{-3}$                  & $64.20 \pm 0.80$            & $65.17\pm0.32$          \\
        \hline\hline\\[-0.25cm]
    \end{tabular}
    \caption{Physical parameters of the \Aur\ system from different studies.}
    \label{tbl:V454_Aur}
\end{table*}

\subsection{Comparison of obtained parameters for \Aur\ system}
\label{txt:dis_comp}

Figure~\ref{fig:our_MLR} shows a comparison of the luminosities, masses, and radii of main-sequence (MS) stars from \citet{2015ASPC..496..164S} with the derived values for both components of the \Aur\ system from this work and the \NN\ system \citep{2020Ap&SS.365..169K}. The comparison of our derived characteristics for the \Aur\ components with published data demonstrates that the properties of \Aur\ fully align with the masses, luminosities, and radii of previously studied stars.

We also independently estimated the distance to \Aur\ and compared it with the distance derived from the latest $Gaia$ satellite data \citep{2016A&A...595A...1G,2023A&A...674A...1G}. The $Gaia$ parallax \citep{2020yCat.1350....0G} translates to a distance of $Dist_{Gaia} = 65.07\pm0.09$ parsecs. We used the $V = 7.65 \pm 0.01$mag value for \Aur\ from the Tycho-2 catalog \citep{2000A&A...355L..27H}, bolometric corrections interpolated from \citet{1981Ap&SS..80..353S}, bolometric luminosities from Table\ref{tab:glob_ELSA}, and extinction $E(B-V)$ from Table~\ref{tab:FBS_stars}. The calculations were performed in Python using the {\tt uncertainties}\footnote{\url{https://pypi.org/project/uncertainties/}} package, which facilitates the computation of output parameter uncertainties while accounting for input parameter errors. The resulting distance, $Dist_{Our} = 65.17\pm0.32$ parsecs, agrees very well with the distance calculated from $Gaia$ data, within the uncertainties. The primary source of error in this estimation is the precision of the photometric data.

As noted earlier in Section~\ref{txt:intro}, two recent papers \citep{2024PARep...2...18Y,2024Obs...144..181S} have been published with a detailed study of the \Aur\ system. Both studies used radial velocity and TESS photometric data, which are also utilized in our work. Table~\ref{tbl:V454_Aur} summarizes the main parameters of the \Aur\ system obtained in all three studies. Since both earlier studies were based on the same spectroscopic data, the mass ratio, and consequently the stellar masses derived, are similar and differ from those obtained in our study. However, this difference does not exceed $2-3\sigma$ of the total error for component A and $1.5-2\sigma$ for component B. Similarly, the derived radii, the size of the semi-major axis ($a$), and the inclination are in comparable agreement. Overall, this is an excellent match, given that all three studies used different software packages for modeling binary systems.

The situation is notably worse for the determined orbital period, where the difference between the value obtained in our work and that from \citet{2024PARep...2...18Y} corresponds to $4.9\sigma$ of the total error. The comparison of the system's eccentricity is even more striking: the difference between our value and that of \citet{2024PARep...2...18Y} amounts to $18\sigma$, while the difference between our value and that of \citet{2024Obs...144..181S} reaches $28\sigma$. This discrepancy may stem from methodological differences, or the reported uncertainties in those studies may be underestimated. The comparison of the temperatures of both system components, which were determined from \`echelle spectra in our study, from photometric data in \citet{2024PARep...2...18Y}, and using the surface brightness ratio in \citet{2024Obs...144..181S}, shows agreement within less than one $\sigma$ of the total error.

Our spectroscopic data allowed us to directly determine the metallicity of \Aur\ as $\mathrm{[Fe/H]} = -0.17\pm0.02$~dex. This value agrees well with the metallicity $\mathrm{[Fe/H]} = -0.14$ reported by \citet{2009A&A...501..941H} and somewhat less so with $\mathrm{[Fe/H]} = -0.08$ determined by \citet{2011A&A...530A.138C} based on spectroscopic and photometric data from the Geneva-Copenhagen Survey \citep{2004A&A...418..989N}. These values are notably different from the metallicity determinations of $\mathrm{[Fe/H]} = -0.02$~dex and $\mathrm{[Fe/H]} = 0.0$~dex reported by \citet{2024PARep...2...18Y} and \citet{2024Obs...144..181S}, respectively, where the authors derived the metallicity of \Aur\ using evolutionary tracks, i.e., by an indirect method.

\begin{figure}
    \centering{
        \includegraphics[clip=,angle=0, width=0.47\textwidth]{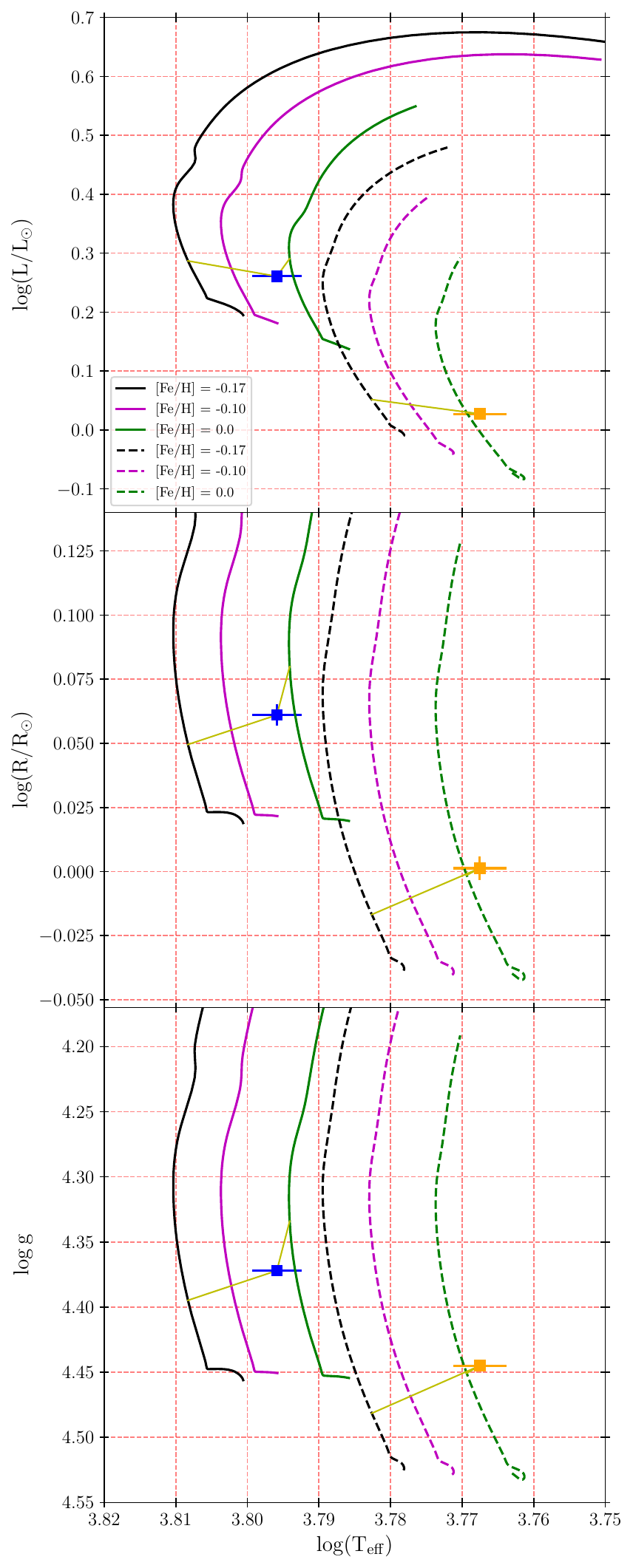}
    }
    \caption{Evolutionary tracks for stars with masses of 1.137~$M_\odot$ (solid lines) and 1.023~$M_\odot$ (dashed lines) for three metallicities: $\mathrm{[Fe/H]} = -0.17$\,dex (black), -0.10 (purple), and 0.00 (green). Only models with $v/v_{crit} = 0$ are shown, as they are indistinguishable from models with $v/v_{crit} = 0.4$. Component A is represented in blue, and component B in orange. The yellow lines indicate the distance to the optimal solution determined using Equation\ref{eq:fit}.       
        \label{fig:V454_evol}}
\end{figure}

\subsection{Is the system \Aur\ synchronised?}
\label{txt:synch}

We began our study of long-period DLEB systems under the assumption that these systems are examples where the components do not influence each other’s evolution, and therefore, the stellar evolution in such systems serves as a true representation of single-star evolution. According to this assumption, we are particularly interested in systems that are neither synchronized nor circularized. The \Aur\ system is certainly not circularized, as it has a significant eccentricity. But what can we say about its synchronization?

As mentioned in Section~\ref{txt:data_analysis}, the \EL\ package includes a parameter responsible for synchronization and, by default, assumes that the stars are synchronized. In our modeling, the value of this parameter was found to be $F = 2.4414 \pm 0.0015$, as derived from Equation~\ref{eqn:synch}, suggesting that the stars in \Aur\ are either already synchronized or close to synchronization.
Attempts to treat the parameter $F$ as a free variable in the modeling did not lead to changes in its value, and adjusting $F$ far from the initial value resulted in a deterioration of the model. Thus, we conclude that our initial assumption of entirely independent evolution for each component in this system appears to be incorrect, as synchronization is a result of tidal interactions between the stars.

This naturally raises the question: how does synchronization affect the evolution of both components, and can we observe its influence in some way?

\subsection{Evolutionary status and age of \Aur}
\label{txt:dis_evol}

Assuming that the studied binary star is a detached system where the components do not influence each other's evolution, the evolutionary status of the components can be assessed based on single-star evolutionary models. In \citet{2019A&A...632A..31G}, it was demonstrated that the results from the \parsec\ \citep[PAdova and TRieste Stellar Evolution Code;][]{2012MNRAS.427..127B}, \basti\ \citep[Bag of Stellar Tracks and Isochrones;][]{2004ApJ...612..168P}, and \mist\ \citep[MESA Isochrones and Stellar Tracks;][]{2016ApJ...823..102C} models are very similar. Therefore, in this work, we used only the \mist\ models. Given that the masses of the \Aur\ components are known with high accuracy, we extracted evolutionary tracks for stars of these masses from the \mist\ database\footnote{\url{http://waps.cfa.harvard.edu/MIST/interp_isos.html}} and analyzed the positions of each component on these tracks for metallicities $\mathrm{[Fe/H]} = -0.17$~dex, $-0.10$, and 0.0. The results are presented in Figure~\ref{fig:V454_evol}, which displays sections of the evolutionary tracks in the coordinates $\log L$--$\log \mathrm{T_{eff}}$, $\log R$--$\log \mathrm{T_{eff}}$, and $\log g$--$\log \mathrm{T_{eff}}$, along with the positions of both components of \Aur. From the figure, it is evident that to match the evolutionary tracks for the given masses and metallicities, both components need to be approximately 200--300~K hotter.
Since we know with certainty that no mass transfer occurred between the components, it is reasonable to hypothesize that this offset results from the influence of synchronization on the evolution of the components. It should be noted that the differences between \mist\ models with $v/v_{crit} = 0$ and $v/v_{crit} = 0.4$ are minimal, and these tracks are indistinguishable.

How observable is the discovered effect in other binary systems similar to \Aur? We examined all DLEBs listed in Table~IV of \citet{2001Obs...121..315G}, which includes 14 binary systems with properties similar to those of \Aur. The \Aur\ system itself is also included in this table. Upon analysis, approximately half of the systems listed have only indirect metallicity estimates and can therefore be excluded from the comparison. All other DLEBs in this table, where metallicity was determined directly using spectroscopy, exhibit, to varying degrees, a similar type of offset toward cooler temperatures relative to the evolutionary track for the given mass and metallicity. Since we hypothesize that this effect may reflect the degree of synchronization within the system, the variability of such an offset is quite logical. Additionally, in our study of the long-period system \NN\ \citep[$P = 99.25$~days;][]{2020Ap&SS.365..169K}, consisting of two F-type stars, we also observed the same temperature offset for both components. In \citet{2020Ap&SS.365..169K}, we also verified that the determined metallicity is independent of the stellar models used in the \fbs\ program.

A comparison of Figure~\ref{fig:V454_evol} with Figure~\ref{fig:our_MLR} suggests that this temperature offset should not result in significant deviations in radius or luminosity. Therefore, we attempted to estimate the age of the system by minimizing the following function:
\begin{equation}
    \label{eq:fit}
    \chi^2 = \sum_{i=1}^2 \left[ \left(\dfrac{\Delta L}{\sigma_L}\right)_i^2 + 
    \left(\dfrac{\Delta T_\mathrm{eff}}{\sigma_\mathrm{T_{eff}}}\right)_i^2 +
    \left(\dfrac{\Delta R}{\sigma_R}\right)_i^2 +
    \left(\dfrac{\Delta g}{\sigma_{g}}\right)_i^2 \right],
\end{equation}
where the summation is over both components ($i = 1,2$), $\Delta$ represents the logarithmic difference between the model and the observed value, and $\sigma$ is also used in logarithmic scale. The search was performed for each model metallicity under the assumption that both components of the system share the same metallicity. For a metallicity of $\mathrm{[Fe/H]} = -0.17$~dex, the age of the \Aur\ system is estimated to be $1.18\pm0.10$\,Gyr, while for $\mathrm{[Fe/H]} = 0.0$\,dex, the age is estimated at $2.77\pm0.30$\,Gyr. In both cases, as shown in Figure\ref{fig:V454_evol}, both components are on the main sequence. Our age estimate agrees very well with the age of $1.19\pm0.09$~Gyr derived by \citet{2024PARep...2...18Y}, where the authors performed evolutionary modeling for the binary system with parameters of \Aur.

\section{Conclusions}
\label{txt:sum}

The long-period eclipsing binary star \Aur\ was studied using spectroscopic data obtained with the \`echelle spectrograph UFES, mounted on the 1.21~m telescope of the Kourovka Astronomical Observatory at Ural Federal University, and photometric data from the TESS satellite. A radial velocity curve was constructed based on 17 spectra obtained between 2021 and 2023, covering the entire phase space of velocity variations for this binary system. Spectral data, radial velocity curves, and photometric data were modeled, and the orbital and absolute parameters of the \Aur\ system components were determined. Using spectroscopic data, the effective temperatures of both components and the system's metallicity were directly estimated. Our modeling based on TESS photometric data suggests that both components in the system are synchronized or close to synchronization. The obtained parameters of the \Aur\ system components were compared with the evolutionary tracks of \mist\ models, and the age and evolutionary status of both components were evaluated. Comparison with model tracks revealed a systematic offset toward cooler temperatures relative to the evolutionary tracks for the given mass and metallicity. It was found that a similar temperature offset exists in a significant number of binary systems with properties similar to \Aur, and it was proposed that this offset is a result of the interaction between the components due to synchronization.

\begin{acknowledgements}
A.\,K. acknowledges support from the National Research Foundation (NRF) of South Africa.
The work of S.\,Yu.\,Gorda was supported by the Ministry of Science and Higher Education of Russia, 
topic No. FEUZ-2023-0019. 

This paper includes data collected by the TESS mission
and obtained from the MAST data archive at the Space Telescope Science Institute (STScI). 
Funding for the TESS mission is provided by the NASA’s Science
Mission Directorate. STScI is operated by the Association of Universities for
Research in Astronomy, Inc., under NASA contract NAS 5–26555. The following resources were used 
in the course of this work: the NASA Astrophysics Data
System; the SIMBAD database operated at CDS, Strasbourg, France; and the
arhiv scientific paper preprint service operated by Cornell University.

This work has made use of data from the European Space Agency (ESA) mission
{\it Gaia} (\url{https://www.cosmos.esa.int/gaia}), processed by the {\it Gaia}
Data Processing and Analysis Consortium (DPAC,
\url{https://www.cosmos.esa.int/web/gaia/dpac/consortium}). Funding for the DPAC
has been provided by national institutions, in particular the institutions
participating in the {\it Gaia} Multilateral Agreement.
\end{acknowledgements}

\bibliographystyle{raa}
\bibliography{LPEB}

\label{lastpage}
\end{document}